\documentclass[a4paper,eqsecnum,english,a4paper,showpacs,amsmath,amssymb,twocolumn]{revtex4}
\usepackage[latin1]{inputenc}
\usepackage{float}
\usepackage{graphicx}
\usepackage{amssymb}

\makeatletter

\usepackage{dcolumn}

\usepackage{bm}

\usepackage{babel}
\makeatother
\begin{document}

\title{Analysis of 
resonant inelastic x-ray scattering 
from Sr$_2$IrO$_4$ in an itinerant-electron approach}

\author{Jun-ichi Igarashi$^{1}$ and Tatsuya Nagao$^{2}$}

\affiliation{
 $^{1}$Faculty of Science, Ibaraki University, Mito, Ibaraki 310-8512,
Japan\\
$^{2}$Faculty of Engineering, Gunma University, Kiryu, Gunma 376-8515,
Japan
}

\date{\today}

\begin{abstract}

We analyze the resonant x-ray scattering (RIXS) spectra from Sr$_2$IrO$_4$ 
in an itinerant electron approach. Employing a multi-orbital tight-binding 
model on the square lattice, we calculate the one-electron energy band 
within the Hartree-Fock approximation, which leads to an antiferromagnetic 
ground state. We then evaluate the two-particle Green's functions for the 
particle-hole pair excitations within the random phase approximation,
which are connected to the RIXS spectra within the fast collision 
approximation. The calculated RIXS spectra exhibit two-peak structure with 
slightly different energies in the low-energy region,
which are originated from the bound states in the two-particle Green's function. 
They may be interpreted as the split modes of magnon.
We also obtain several $\delta$-function peaks, which arise from the bound 
states around the bottom of energy continuum. They may be called as the 
exciton modes. These characteristics are in qualitative agreement with the 
RIXS experiment, demonstrating that the weak coupling theory could explain 
both the magnon and the exciton peaks in the RIXS spectra on an equal footing.

\end{abstract}


\maketitle

\section{\label{sect.1}Introduction}

Strong synchrotron sources have been rapidly developing,
and resonant inelastic x-ray scattering (RIXS) has become a powerful tool 
to probe elementary excitations in solids \cite{Ament2011-rmp,Ishii2013}.
Both the $K$- and $L$-edge resonances are utilized in transition-metal 
compounds.
On the $K$ edge resonance, the $1s$ core-electron is prompted to empty 
$p$-symmetric states by absorbing photon, then the photo-excited electron is 
recombined with the core hole by emitting photon. 
In this process, charge excitations are created to screen the core-hole 
potential in the intermediate state. Note that two magnons could also be 
created in magnetic systems \cite{Nagao2007,vdBrink2007,Hill2008,Forte2008}, 
since the exchange coupling is modified around 
the core-hole site in the presence of the core-hole potential.

On the $L$-edge resonance, on the other hand, the $2p$ core-electron is 
prompted to empty $d$-symmetric states by absorbing photon, then an 
electron occupied on the $d$-symmetric state is combined with the core hole 
by emitting photon. Since the electron combined with the core hole is not 
necessarily the same as the photo-excited electron, 
the particle-hole pair excitations could be
directly created in the $d$-symmetric states in addition to the screening
effect of the core-hole potential in the intermediate state.  
Note that the $2p$-core states are split into two well-separated 
levels with the total angular momentum $j_c=1/2$ and $j_c=3/2$ due to the 
strong spin-orbit interaction (SOI).
The corresponding $L$-edges are discriminated as the
$L_2$ and $L_3$-edges, respectively.
Owing to this split, the single spin-flip excitations could be created.
Actually, the spectral peaks 
as a function of energy loss are found to follow the 
dispersion of spin waves in the Heisenberg model with changing momentum 
transfer in undoped 
cuprates \cite{Braicovich2009,Braicovich2010,Guarise2010}.
Such $L$-edge RIXS spectra have been analyzed on the spin model
within the fast collision approximation (FCA),
which is justified when the core-hole life-time broadening width is larger 
than the concerned excitation 
energy \cite{Ament2007,Ament2009,Haverkort2010}.
Although only the one-magnon excitations could be created within the FCA, 
the experimental energy profile shows asymmetric 
shape \cite{Braicovich2009,Braicovich2010,Guarise2010},
indicating that two-magnon excitations are involved in addition to the 
one-magnon excitations. An analysis going beyond the FCA has been carried out
to explain the asymmetrical profile in quantitative agreement with the 
experiment \cite{Igarashi2012-1,Igarashi2012-2,Nagao2012}. 

Recently, RIXS experiments have been carried out at the Ir L$_3$ edge
in Sr$_2$IrO$_4$ \cite{Ishii2011,J.Kim2012}.
This material shows the antiferromagnetism at low 
temperatures below $\sim$ 230 K \cite{Crawford1994,Cao1998,Moon2006}.
Its ordering is concerned with the spin-orbital 
coupled \emph{isospin} $j_{\rm eff}=1/2$, since the SOI is large on 
the $5d$ states of Ir \cite{Jackeli2009}. 
The low-energy peak behaves like the 
one-magnon peak in undoped cuprates \cite{J.Kim2012}, while other peaks 
emerge around $\omega\sim 0.5$ eV 
with substantial weights as a function of energy loss $\omega$.
The RIXS spectra have been analyzed within the FCA
on the basis of the localized electron 
model \cite{Ament2011,Igarashi2014-1}.
Note that the FCA is expected to work well for this material,
since the $2p$-core hole has the life-time broadening width
as large as $2.5-3.5$ eV \cite{Krause1979}, 
which is much larger than the concerned excitation 
energies. The low-energy peak has been interpreted as the magnetic excitation
(magnon) in the isospin space $j_{\rm eff}=1/2$, while the peak around
$\omega\sim 0.5$ eV as the excitation (exciton) from the $j_{\rm eff}=3/2$
manifold to $j_{\rm eff}=1/2$ manifold\cite{Katukuri2012}. 
Recently, the magnon mode has been 
predicted to be split into two modes due to the interplay
between Hund's coupling and the SOI, and a detailed analysis of the
magnon peak has been made with taking account of the mode 
splitting \cite{Igarashi2013-2,Igarashi2014-1}. 

Although the localized electron model has been successful in analyzing
the RIXS spectra, there remain issues that the itineracy of the electrons
might play a key role to elucidate the physical features of
Sr$_2$IrO$_4$. For instance, 
it has been argued whether the system behaves like the Mott insulator 
or the band insulator \cite{Arita2012,Moser2014}. 
It is also established that the isotropic Heisenberg model can reproduce
the magnon dispersion only when it includes the second and third
nearest neighbor exchange coupling in addition to the first
nearest neighbor exchange coupling \cite{Kim2012}.
Such observations naturally prompt us to investigate the material
in the itinerant electron approach and 
several attempts have been carried out to study its electronic structure. 
The band structure calculation has been carried out within the local 
density approximation augmented by the Coulomb interaction (LDA+U),
having led to the antiferromagnetic ordering and the associated energy 
gap in the one-electron energy band \cite{Kim2008}. 
The electron correlation effects have been taken into account by the 
variational method \cite{Watanabe2010}, 
as well as by the dynamical mean field theory \cite{Arita2012}.
Recently, excitation spectra have been investigated by
calculating the spectral function of the two-particle Green's function
for particle-hole pair excitations within the Hartree-Fock 
approximation (HFA) and the random phase approximation 
(RPA) \cite{Igarashi2014-2}, having led that they are 
composed of magnons and excitons.
However, a direct comparison between the spectral function and
the RIXS spectrum cannot be allowed, 
since the former is different from the RIXS spectra due to the 
second-order optical process. 

The purpose of this paper is to analyze the RIXS spectra 
with an argument based on an itinerant electron picture.
Introducing the multi-orbital tight-binding model, 
we calculate the one-electron energy band within the HFA, where
the antiferromagnetic ground state is realized. 
Then, on this ground state, 
we calculate the two-particle Green's function for the particle-hole pair 
excitations within the RPA. 
Magnons appear as bound states below the energy continuum, and are split into
two modes. Other several bound states emerge around the bottom of the energy 
continuum \cite{Com3}. These together with the continuum states 
(often containing resonant modes) may constitute exciton modes. 

The spectral function of the two-particle Green's function 
are related with the RIXS spectra within the FCA, which is known to work well
for the RIXS spectra in Sr$_2$IrO$_4$.
For magnons split into two modes, the two-peak structures are found with 
significant momentum dependence of intensities. 
The splitting of the magnon modes, however, has not been
confirmed yet by RIXS experiments, probably because the experimental 
energy resolution is as large as $40$ meV. 
Furthermore, sharp exciton peaks emerge with their intensities 
larger than the magnon intensities, being separated from the magnon peaks, 
in consistent with the RIXS experiment.  
Thus it is demonstrated that the weak coupling approach of the HFA and RPA 
could explain both the magnon and the exciton peaks in the RIXS spectra
on an equal footing, providing a good starting point of taking account of
electron correlations.

The present paper is organized as follows. In Sec. \ref{sect.2},
we introduce a multi-orbital tight-binding model, and study the electronic 
structure within the HFA. 
In Sec. \ref{sect.3}, we describe the dipole process, and calculate the
absorption coefficient at the $L$ edge.
In Sec. \ref{sect.4}, we derive the formula for the RIXS spectra
within the FCA.
In Sec. \ref{sect.5}, comparisons are made between the calculated and
experimental RIXS spectra.
Section \ref{sect.6} is devoted to the concluding remarks. 

\section{\label{sect.2}Electronic structure}
\subsection{Model Hamiltonian}
Transition metal oxide Sr$_2$IrO$_4$ with 
the K$_2$NiF$_4$-type structure
is composed of IrO$_2$ layer separated by Sr-O layer \cite{Crawford1994}. 
Since the crystal field energy of the $e_g$ orbitals is about 2 eV higher 
than that of the $t_{2g}$ orbitals, 
we consider only $t_{2g}$ orbitals with five electrons occupying
in each Ir atom.
Since the oxygen octahedra surrounding an Ir atom are rotated 
about the crystallographic $c$ axis by 
about 11$^{\circ}$ \cite{Watanabe2010,Wang2011},
the $t_{2g}$ states are defined in the local coordinate frames rotated 
in accordance with the rotation of octahedra.
For simplicity, disregarded is the fact that the degenerate $t_{2g}$ levels are 
split by the tetragonal crystal field due to the rotation and distortion 
of IrO$_6$ octahedra. Then, 
we start from the multi-orbital Hubbard model
on the square lattice in the local coordinate frames, 
\begin{equation}
 H = H_{\rm kin} + H_{\rm SO}+H_{\rm I},
\end{equation}
where $H_{\rm kin}$, $H_{\rm SO}$, and $H_{\rm I}$ are
described by 
the annihilation ($d_{in\sigma}$) and creation 
($d_{in\sigma}^{\dagger}$) operators of 
an electron with orbital $n$ ($=yz,zx,xy$) and spin $\sigma$ at the Ir site $i$ as follows.
\begin{eqnarray}
H_{\rm kin} & = & \sum_{\left\langle i,i'\right\rangle }
\sum_{n,n',\sigma}t_{in,i'n'}d_{in\sigma}^{\dagger}d_{i'n'\sigma}+ {\rm H.c.},
\\
H_{\rm SO} & = & \zeta_{\rm SO}\sum_{i}\sum_{n,n',\sigma,\sigma'}
 d_{in\sigma}^{\dagger}({\bf L})_{nn'}
 \cdot({\bf S})_{\sigma\sigma'}d_{in'\sigma'}, \\
H_{\rm I} & = & 
  U\sum_{i,n} n_{in\uparrow}n_{in\downarrow} \nonumber \\
  &+&
 \sum_{i,n<n',\sigma}[U' n_{in\sigma}n_{in'-\sigma}
                 + (U'-J) n_{in\sigma}n_{in'\sigma}] \nonumber\\
 &+&J\sum_{i,n\neq n'} (d_{in\uparrow}^{\dagger}d_{in'\downarrow}^{\dagger}
                     d_{in\downarrow}d_{in'\uparrow}
                    +d_{in\uparrow}^{\dagger}d_{in\downarrow}^{\dagger}
                     d_{in'\downarrow}d_{in'\uparrow}). \nonumber\\ 
\end{eqnarray}
The inter-site interaction $H_{\rm kin}$ stands for
the kinetic energy.
The transfer integral $t_{in,i'n'}$ exhibits a highly anisotropic 
nature. An electron on the $xy$ orbital could transfer to the $xy$ 
orbital in the nearest neighbor sites through the intervening O $2p$ orbitals,
while an electron on the $yz$($zx$) orbital could transfer to 
the $yz$($zx$) orbital in the nearest-neighbor sites only 
along the $y$($x$) direction. 
The none-zero values of $t_{in,i'n'}$'s are assumed to be the 
same and denoted as $t_1$.
The SOI of $5d$ electrons is denoted as $H_{\rm SO}$ 
with ${\bf L}$ and ${\bf S}$ denoting the orbital and spin angular momentum 
operators.
The $H_{\rm I}$ represents the Coulomb interaction between electrons with
$\nu=(n,\sigma)$. Parameters satisfy $U=U'+2J$ \cite{Kanamori1963}.
We use the values $U=1.4$ eV, and $J/U=0.15$ in the following calculation. 
As regards the transfer integral $t_1$ and the SOI parameter $\zeta_{\rm SO}$, 
we consider two typical parameter sets; one is that $\zeta_{\rm SO}=0.36$ eV, 
$t_1=0.36$ eV (Case A), and another is that $\zeta_{\rm SO}=0.45$ eV, 
$t_1=0.25$ eV (Case B). The values in Case A are the same as in
[\onlinecite{Watanabe2013}], and give the one-electron band width
consistent with the band calculation based on the local density 
approximation\cite{Kim2008,Igarashi2014-2}. The smaller value of $t_1$
in Case B may lead to the larger energy gap in the one-electron band. 
Note that the smaller values of $t_1$ around 0.1-0.2 eV have been
estimated on the basis of a localized picture\cite{Perkins2014}.

\subsection{Hartree-Fock Approximation}
A unit cell $j$ contains two atoms at ${\bf r}_j$ and at 
${\bf r}_j+{\bf a}$, where ${\bf a}=(a,0)$ with $a$ a nearest neighbor 
distance. 
We introduce the Fourier transform of annihilation operator 
with the wave vector ${\bf k}$
in the magnetic Brillouin zone (MBZ), which is defined 
as the half of the first Brillouin zone:
\begin{equation}
 d_{\lambda n\sigma}({\bf k}) = \sqrt{\frac{2}{N}}\sum_{j}d_{j'n\sigma}
                             \textrm{e}^{-i{\bf k}\cdot{\bf r}_j} ,
\label{eq.Fourier}
\end{equation}
where $j$ runs over unit cells and $N/2$ stands for the number of the
unit cells. We assign $\lambda=1$ and $2$ for the A 
and B sublattices, respectively. The index $j'$ specifies the site
within the $j$-th unit cell as $\textbf{r}_{j'}={\bf r}_j$ and 
${\bf r}_j+{\bf a}$ for $\lambda =1$ and $2$, respectively.
Then, the one-electron energy $H_0\equiv H_{\rm kin}+H_{\rm SO}$
may be rewritten as
\begin{equation}
 H_{0} = \sum_{{\bf k}\xi\xi'} d_{\xi}^{\dagger}({\bf k})
             \left[\hat{H}_0({\bf k})\right]_{\xi,\xi'}d_{\xi'}({\bf k}),
\end{equation}
with abbreviations $\xi=(\lambda,n,\sigma)$ and $\xi'=(\lambda',n',\sigma')$.

Arranging $\xi$ in order 
$(1,yz,\uparrow)$, $(1,zx,\uparrow)$, $(1,xy,\uparrow)$,
$(1,yz,\downarrow)$, $(1,zx,\downarrow)$, $(1,xy,\downarrow)$,
$(2,yz,\uparrow)$, $(2,zx,\uparrow)$, $(2,xy,\uparrow)$,
$(2,yz,\downarrow)$, $(2,zx,\downarrow)$, $(2,xy,\downarrow)$,
we have $\hat{H}_0({\bf k})$ in a block form,
\begin{equation}
 \hat{H}_{0}({\bf k}) = \left(
    \begin{array}{cc}
     \hat{H}_{\rm AA}^{0}({\bf k}) & \hat{H}_{\rm AB}^{0}({\bf k}) \\ 
     \hat{H}_{\rm BA}^{0}({\bf k}) 
                                   & \hat{H}_{\rm BB}^{0}({\bf k})  
    \end{array} \right),
\end{equation}
where
\begin{eqnarray}
 && \hat{H}_{\rm AA}^{0}({\bf k}) = 
  \hat{H}_{\rm BB}^{0}({\bf k}) = \frac{\zeta_{\rm SO}}{2}\left(
     \begin{array}{cccccc}
     0 & i & 0 & 0 & 0 & -1 \\
     -i & 0 & 0 & 0 & 0 & i \\
     0 & 0 & 0 & 1 & -i & 0 \\
     0 & 0 & 1 & 0 & -i & 0 \\
     0 & 0 & i & i & 0 & 0 \\
     -1 & -i & 0 & 0 & 0 & 0 
     \end{array}
     \right), \nonumber \\
\\
 && \hat{H}_{\rm AB}^{0}({\bf k}) = 
  [\hat{H}_{\rm BA}^{0}({\bf k})]^{\star} \nonumber \\
&&= \left(
     \begin{array}{cccccc}
     t_1({\bf k}) & 0 & 0 & 0 & 0 & 0 \\
     0 & t_2({\bf k}) & 0 & 0 & 0 & 0 \\
     0 & 0 & t_3({\bf k}) & 0 & 0 & 0 \\
     0 & 0 & 0 & t_1({\bf k}) & 0 & 0 \\
     0 & 0 & 0 & 0 & t_2({\bf k}) & 0 \\
     0 & 0 & 0 & 0 & 0 & t_3({\bf k}) 
     \end{array}
     \right).
\end{eqnarray}
Here the dispersion may be expressed as
\begin{equation}
t_{n} (\textbf{k}) = - 2 t_1 {\rm e}^{-ik_x}
\times \left\{ \begin{array}{lcl}
\cos k_y, & \textrm{for} & n=1 \\
\cos k_x, & \textrm{for} & n=2 \\
(\cos k_x+\cos k_y), & \textrm{for} & n=3 \\
\end{array} \right. ,
\end{equation}
where 
${\bf k}$ is measured in units of $1/a$. 

We follow the conventional procedure of the HFA as explained in 
Ref. \onlinecite{Igarashi2013-1}.
Rewriting $H_{\rm I}  =  \frac{1}{2}\sum_{i}\sum_{\nu_1,\nu_2,\nu_3,\nu_4}
 g(\nu_1\nu_2;\nu_3\nu_4) d_{i\nu_1}^{\dagger}d_{i\nu_2}^{\dagger}
 d_{i\nu_4}d_{i\nu_3}$, we replace $H_{\rm I}$ by
\begin{equation}
 H_{\rm I}^{\rm HF} = \frac{1}{2}\sum_{j}\sum_{\xi_1,\xi_2,\xi_3,\xi_4}
 \Gamma^{(0)}(\xi_1\xi_2;\xi_3\xi_4)   
  \langle d_{j\xi_2}^{\dagger}d_{j\xi_4}\rangle d_{j\xi_1}d_{j\xi_3},
\end{equation}
where $\Gamma^{(0)}$ is the antisymmetric vertex function,
\begin{equation}
 \Gamma^{(0)}(\xi_1\xi_2;\xi_3\xi_4)=
  g(\xi_1\xi_2;\xi_3\xi_4)-g(\xi_1\xi_2;\xi_4\xi_3),
\end{equation}
with $\xi=(\lambda,\nu)$.
Here, $d_{j \xi}=d_{j (\lambda \nu)}$ denotes the annihilation
operator of the $d$ electron with $\nu$ spin-orbital state
at the site belonging to the sublattice $\lambda$ in the $j$-th
unit cell.
Then, we introduce the single-particle Green's function 
in a matrix form with $12\times 12$ dimensions,
\begin{equation}
\left[\hat{G}({\bf k},\omega)\right]_{\xi,\xi'}=-i\int\langle 
 T[d_{\xi}({\bf k},t)d_{\xi'}^{\dagger}({\bf k},0)]\rangle
 {\rm e}^{i\omega t}{\rm d}t,
\label{eq.dG}
\end{equation}
where $T$ is the time ordering operator, and $\langle X \rangle$
denotes the ground-state average of operator $X$.
The Green's function is obtained by solving the equations of motion,
resulting in 
\begin{equation}
[\hat{G}({\bf k},\omega)]_{\xi,\xi '}
=\sum_{\ell}
\frac{[\hat{U}({\bf k})]_{\xi,\ell}[\hat{U}({\bf k})^{-1}]_{\ell,\xi '}}
{\omega-E_{\ell}({\bf k})+i\epsilon{\rm sgn}[E_{\ell}({\bf k})]}, 
\label{eq.Green}
\end{equation}
where ${\rm sgn}[A]$ stands for a sign of quantity $A$ and 
$\delta$ denotes a positive convergent factor.
The $\ell$-th energy eigenvalue within the HFA is written as 
$E_{\ell}({\bf k})$ measured from the chemical potential. 
The definition of the unitary matrix $\hat{U}({\bf k})$ is found in Ref. 
\onlinecite{Igarashi2013-1}.
The Green's function contains the expectation values of the electron density 
operator on the ground state, which are self-consistently determined from
\begin{equation}
\langle a_{\xi}^{\dagger}a_{\xi'}\rangle=
\frac{2}{N}\sum_{{\bf k}}
\int [-i \hat{G}({\bf k},\omega)]_{\xi,\xi'}
{\rm e}^{i\omega0^{+}}\frac{{\rm d}\omega}{2\pi}. 
\label{eq.gap}
\end{equation}
This equation is solved by iteration with summing over ${\bf k}$ by dividing 
the MBZ into $100\times 100$ meshes.
We obtain a self-consistent solution of the 
antiferromagnetic order with the staggered magnetic moment along 
the $x$ axis as the ground state, which is consistent with the magnetic 
measurements \cite{Crawford1994,Cao1998}.
Both the orbital and the spin moments are induced due to the strong SOI; 
$\langle S_x\rangle= \pm 0.112$ (Case A) and $\pm 0.143$ (Case B), while
$\langle L_x\rangle= \pm 0.435$ (Case A) and $\pm 0.551$ (Case B).
The antiferromagnetic order in the local coordinate frames 
indicates that a weak ferromagnetic moment is induced in the  
coordinate frame fixed to the crystal axes. 
The one-electron energy has a finite gap due to the antiferromgnatic
order.

\section{\label{sect.3}Dipole transition and absorption spectra}

\subsection{Dipole transition}
For the interaction between photon and matter, we consider the dipole
transition at the $L$ edge, where the $2p$ core-electron is excited
to the $5d$ states by absorbing photon (and the reverse process).
This process may be described by the interaction 
\begin{eqnarray}
H_{x}&=&\sum_{\lambda,n,\sigma,j_{c}, m,\alpha}
w(n\sigma;j_{c}m;\alpha) \nonumber \\
&\times&\sum_{\bf{k},\bf{q}}
d_{\lambda n\sigma}^{\dagger}({\bf k+q})p_{\lambda j_{c}m}({\bf k})
c_{\alpha}({\bf q})v(\lambda,{\bf q})+{\rm H.c.}, \nonumber \\
\end{eqnarray}
where $c_{\alpha}({\bf q})$ is the annihilation operator of photon 
with momentum ${\bf q}$ and polarization $\alpha$. The
$p_{\lambda j_{c}m}({\bf k})$ is the annihilation operator of 
core electron belonging to the $\lambda$ site
with wave vector ${\bf k}$ and the angular momentum $j_c=3/2$ and $1/2$
and magnetic quantum number $m$. 
The $v(\lambda,{\bf q})$ stands for the extra phase on the B sites,
which is explicitly defined as
\begin{equation}
v(\lambda,{\bf q})=\delta_{\lambda,1}+\delta_{\lambda,2}{\rm e}^{iq_x}.
\end{equation}
Note that, when ${\bf k+q}$ lies outside the first MBZ, it is reduced back
to the inside of the first MBZ by a reciprocal lattice vector in the
reduced zone scheme.  The $w(n\sigma;j_{c}m;\alpha)$ represents the matrix 
elements of the $2p\to 5d$ transition. Table I lists the values for 
$\alpha=x$, $y$, $z$ corresponding to the polarization directing to the 
$x$, $y$, $z$ axes.

\begin{table}
\caption{$w(n\sigma;j_{c} m;\alpha)$ within $t_{2g}$ basis with 
$j_{c}=\frac{3}{2}$ and $\frac{1}{2}$.\label{table.w}
}
\begin{ruledtabular}
\begin{tabular}{cc|cccc|cc}
& & \multicolumn{4}{c|} {$j_{c}=\frac{3}{2}$} & 
\multicolumn{2}{c}{$j_{c}=\frac{1}{2}$} \\
$\alpha$ &$(n\sigma) \setminus m$ & $\frac{3}{2}$ & $\frac{1}{2}$ & $-\frac{1}{2}$ & $-\frac{3}{2}$ &
$\frac{1}{2}$ & $-\frac{1}{2}$ \\
\hline
$x$& $(zx \uparrow)$ & 0 & $\frac{\sqrt{2}}{\sqrt{15}}$ & 0 & 0 & 
$\frac{1}{\sqrt{15}}$ & 0 \\
 & $(zx \downarrow)$ & 0 & 0 & $\frac{\sqrt{2}}{\sqrt{15}}$ & 0 &
0 & $-\frac{1}{\sqrt{15}}$ \\
 & $(xy \uparrow)$ & $-\frac{i}{\sqrt{10}}$ & 0 & $-\frac{i}{\sqrt{30}}$ & 0 &
0 & $-\frac{i}{\sqrt{15}}$\\
 & $(xy \downarrow)$ & 0 & $-\frac{i}{\sqrt{30}}$ & 0 & $-\frac{i}{\sqrt{10}}$ &
$\frac{i}{\sqrt{15}}$ & 0 \\
\hline
$y$ & $(yz \uparrow)$ & 0 & $\frac{\sqrt{2}}{\sqrt{15}}$ & 0 & 0 &
$\frac{1}{\sqrt{15}}$ & 0 \\
 & $(yz \downarrow)$ & 0 & 0 & $\frac{\sqrt{2}}{\sqrt{15}}$ & 0 &
0 & $-\frac{1}{\sqrt{15}}$ \\
 & $(xy \uparrow)$ & $-\frac{1}{\sqrt{10}}$ & 0 & $\frac{1}{\sqrt{30}}$ & 0 &
0 & $\frac{1}{\sqrt{15}}$  \\
 & $(xy \downarrow)$ & 0 & $-\frac{1}{\sqrt{30}}$ & 0 & $\frac{1}{\sqrt{10}}$ &
$\frac{1}{\sqrt{15}}$ & 0 \\
\hline
$z$ & $(yz \uparrow)$ & $-\frac{i}{\sqrt{10}}$ & 0 & $-\frac{i}{\sqrt{30}}$ & 0 &
0 & $-\frac{i}{\sqrt{15}}$ \\
 & $(yz \downarrow)$ & 0 & $-\frac{i}{\sqrt{30}}$ & 0 & $-\frac{i}{\sqrt{10}}$ &
$\frac{i}{\sqrt{15}}$ & 0 \\
 & $(zx \uparrow)$ & $-\frac{1}{\sqrt{10}}$ & 0 & $\frac{1}{\sqrt{30}}$ & 0&
0 & $\frac{1}{\sqrt{15}}$  \\
 & $(zx \downarrow)$ & 0 & $-\frac{1}{\sqrt{30}}$ & 0 & $\frac{1}{\sqrt{10}}$ &
$\frac{1}{\sqrt{15}}$ & 0\\
\end{tabular}
\end{ruledtabular}
\end{table}

\subsection{Absorption coefficient at the $L$ edge}
X ray could be absorbed by exciting the $2p$ electron to 
unoccupied levels at the $L$ edge. Since the core states are well localized
in real space, the absorption coefficient is given by summing the intensity 
at each sites. Neglecting the interaction between the excited electron and
the core hole left behind, we have the expression of the absorption coefficient as
\begin{eqnarray}
 A(\omega_i,j_c) &\propto& \frac{2}{N}\sum_{\alpha,m}\sum_{{\bf k},\ell}
\sum_{\xi,\xi'}
 w(n\sigma;j_c m;\alpha)w^*(n'\sigma':j_c m;\alpha)
  \nonumber\\
 &\times& 
 \frac{\delta_{\lambda,\lambda '} 
  U^{*}_{\xi,\ell}({\bf k})U_{\xi',\ell}({\bf k})[1-n_{\ell}({\bf k})]}
      {[\omega_i-E_{\ell}({\bf k})+\epsilon_{\rm 2p}(j_c)]^2+\Gamma_c^2},
\label{eq.abs}
\end{eqnarray}
where $\xi=(\lambda,n,\sigma)$ and $\xi'=(\lambda ',n',\sigma')$.
The occupation number of the eigenstate
with energy $E_{\ell}({\bf k})$ is given by $n_{\ell}({\bf k})$.
The $\omega_i$ and $\epsilon_{\rm 2p}(j_c)$ 
represent the energies of the incident photon 
and of the $2p$ core-hole level in the $j_c$ manifold, respectively.
The lifetime broadening width of the core-hole is given 
by $\Gamma_c$. Note that polarizations are averaged over
in Eq. (\ref{eq.abs}). 

\begin{figure}
\includegraphics[width=8.0cm]{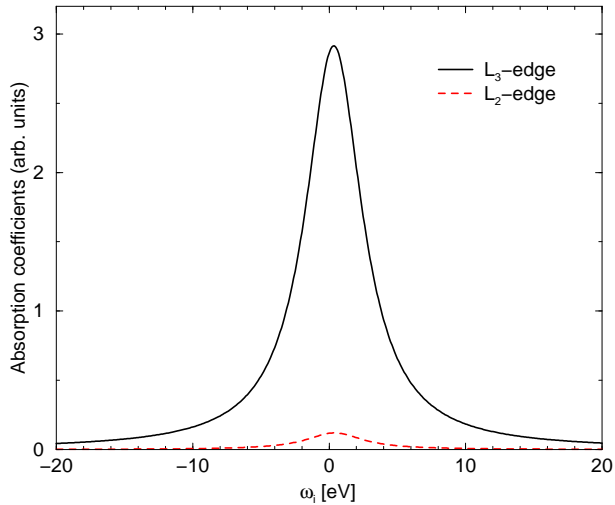}%
\caption{\label{fig.abs} (Color online)
Absorption coefficients $A(\omega_i,j_c)$ as a function of x-ray energy 
with $\Gamma_c=2.5$ eV in Case A. The (black) solid and the (red) broken lines
are spectra at the Ir $L_{3}$ and $L_{2}$ edges, respectively. 
The origin of energy is set to be the difference 
between the bottom of the conduction band and the core-level energy. 
The absorption coefficients for the parameters in Case B are almost the same 
as in Case A.
}
\end{figure}

Figure \ref{fig.abs} shows the calculated absorption coefficient with 
$\Gamma_c=2.5$ eV for the parameters in Case A. The origin of energy is set 
to be the difference between the bottom of the conduction band and each 
core-level energy. Since the conduction band width is of order 1 eV, 
which is smaller than $\Gamma_c$,
the spectral shape looks quite similar to the Lorentzian shape. Note that 
the interaction between the $2p$ core-hole and the excited electron would 
make the spectral shape more sharper.
The intensity at the $L_2$ edge is found much smaller than that at the 
$L_3$ edge in agreement with the experiment and the analysis 
with the localized states in the 
$j_{\rm eff}=\frac{1}{2}$ manifold \cite{Kim2009,Clancy2012}.
The present result accordingly indicates that the conduction band given by 
the HFA is mainly composed of the states in the $j_{\rm eff}=\frac{1}{2}$ 
manifold. The absorption coefficients for the parameters in Case B are almost 
the same as in Case A. 

\section{\label{sect.4}Formula for RIXS spectra}
\subsection{Second-order optical process}
The RIXS spectral intensity may be expressed as the second-order 
optical process,
\begin{eqnarray}
 W(\omega_i,q;\alpha_i,\alpha_f) &=& 2\pi\sum_{f}
 \left | \sum_{n}
 \frac{\langle\Phi_f|H_x|\Phi_n\rangle\langle\Phi_n|H_x|\Phi_i\rangle}
 {\omega_i+\epsilon_g-\epsilon_n+i\Gamma_c}\right|^2 \nonumber \\
 &\times& \delta(\omega_i+\epsilon_g-\omega_f-\epsilon_f) .
\label{eq.second}
\end{eqnarray}
The initial state is given by
$|\Phi_i\rangle=c_{\alpha_i}^{\dagger}({\bf q}_i)|0\rangle|g\rangle$, where
$|g\rangle$ represents the ground state of the matter with energy 
$\epsilon_g$, and $|0\rangle$ denotes the vacuum state
with photon. The intermediate state is given by 
$|\Phi_n\rangle=|0\rangle|n\rangle$, where
$|n\rangle$ stands for the intermediate state of the matter with energy 
$\epsilon_n$. 
The final state is given by
$|\Phi_f\rangle=c_{\alpha_f}^{\dagger}({\bf q}_f)|0\rangle|f\rangle$, 
where $|f\rangle$ represents the final state of the matter with energy 
$\epsilon_f$. 
The incident photon has momentum and energy $q_i=({\bf q}_i,\omega_i)$, 
and polarization $\alpha_i$, while the scattered photon has momentum 
and energy $q_f=({\bf q}_f,\omega_f)$, and polarization $\alpha_f$. 
The momentum and energy transferred to the matter are accordingly given
by $q=q_i-q_f=({\bf q},\omega)$.

In this second-order process, the dipole transition creates the 
$(5d)^6$-configuration at the core-hole site in the intermediate state. 
This state would be relaxed by hopping the excited electron 
to neighboring sites. Since the conduction band has the width of
at most 1 eV while $\Gamma_c$ is as large as 2.5 eV,
the energy denominator of Eq.~(\ref{eq.second}) could be factored out
in a reasonable accuracy. It may be hard to create additionally electron-hole 
pairs in the intermediate state, since the $(5d)^6$-configuration 
is almost kept at the core-hole site.
Therefore, Eq.~(\ref{eq.second}) may be approximated as
\begin{eqnarray}
&& W(\omega_i,q;\alpha_i,\alpha_f) 
\nonumber\\
&=& 2\pi
  |R(\omega_i,E_0)|^2\sum_f 
 \left|\sum_n
    \langle \Phi_f|H_x|\Phi_n\rangle\langle\Phi_n|H_x|\Phi_i\rangle\right|^2 \nonumber \\
&\times&  \delta(\omega+\epsilon_g-\epsilon_f),
\end{eqnarray}
where
\begin{equation}
 R(\omega_i,E_0)=\frac{1}{\omega_i-E_0+\epsilon_{\rm 2p}(j_c)
                         +i\Gamma_c}. \label{eq.R}
\end{equation}
Equation (\ref{eq.R}) arises from the energy denominator
factored out with $E_0$ being a typical energy of the conduction band.
Moreover, the intensity is rewritten as
\begin{eqnarray}
W(\omega_i,q;\alpha_i,\alpha_f)
 &=& |R(\omega_i,E_0)|^2
 \hat{M}^{\dagger}({\bf q},\alpha_i,\alpha_f;j_c) \nonumber \\
&\times&\hat{Y}^{+-}(q)
 \hat{M}({\bf q},\alpha_i,\alpha_f;j_c),
\label{eq.second2}
\end{eqnarray} 
where
\begin{equation}
\left[\hat{Y}^{+-}({\bf q},\omega)\right]_{\xi_1\xi'_{1};\xi\xi'}
=  
\int_{-\infty}^{\infty} 
\langle [\rho_{{\bf q}\xi_1\xi'_1}(t)]^{\dagger}
       \rho_{{\bf q}\xi\xi'}(0)\rangle {\rm e}^{i\omega t}{\rm d}t,
\end{equation}
with
\begin{equation}
  \rho_{{\bf q}\xi\xi'} = \sqrt{\frac{2}{N}}\sum_{\bf k}
   d_{\xi}^{\dagger}({\bf k+q})d_{\xi'}({\bf k}).
\end{equation}
Here ${\bf k+q}$ is to be reduced back to the MBZ by a reciprocal 
vector ${\bf G}$, when it lies outside the MBZ.
The $\hat{Y}^{+-}(q)$ represents the correlation function of 
the electron-hole pair excitations, which is a matrix of $144\times 144$ 
dimensions. The $\hat{M}({\bf q},\alpha_i,\alpha_f;j_c)$ is regarded as a 
vector with 144 dimensions, defined by
\begin{eqnarray}
&& \left[\hat{M}({\bf q},\alpha_i,\alpha_f;j_c)\right]_{\xi\xi'}
\nonumber \\
& =& \delta_{\lambda,\lambda'} v(\lambda,{\bf q})  
\sum_{m}\sum_{\alpha,\beta=x,y,z}(\alpha_i)_{\alpha} 
 w(n\sigma;j_{c}m;\alpha)\nonumber \\
&& \times w^*(n'\sigma';j_{c}m;\beta)
(\alpha_f)_{\beta}, 
\label{eq.Mdef}
\end{eqnarray}
with $\xi=(\lambda,n,\sigma)$ and $\xi'=(\lambda',n',\sigma')$.
Since the scattering event takes place within a single site,
we have the second factor from
\begin{equation}
 v(\lambda,{\bf q}_i)v^{*}(\lambda,{\bf q}_f)=v(\lambda,{\bf q}_i-{\bf q}_f).
\end{equation}
Note that the ${\bf q}$-dependence of $\hat{M}$ does not have the 
periodicity with the MBZ, leading to the RIXS intensities different
between inside and outside the first MBZ.

\subsection{Correlation function within the RPA}
To evaluate the correlation function, it is convenient to introduce the 
time-ordered Green's function,
\begin{equation}
\left[\hat{Y}^{{\rm T}}(q)\right]_{\xi_1\xi'_{1};\xi\xi'}
=-i\int \left\langle 
T\left\{[\rho_{{\bf q}\xi_1\xi'_{1}}(t)]^{\dagger}
\rho_{{\bf q}\xi\xi'}(0)\right\}\right\rangle
{\rm e}^{iq_{0}t}{\rm d}t.
\label{eq.green_time}
\end{equation}
The correlation function is evaluated from the Green's function by
applying the fluctuation-dissipation theorem 
for $\omega>0$ \cite{Igarashi2013-1},
\begin{equation}
 \left[\hat{Y}^{+-}(q)\right]_{\xi_1\xi'_{1};\xi\xi'}=
 -i\left\{\left[\hat{Y}^{{\rm T}}(q)
 \right]^{*}_{\xi\xi';\xi_1\xi'_{1}}
        -\left[\hat{Y}^{{\rm T}}(q)\right]_{\xi_1\xi'_{1};\xi\xi'}
\right\}.
\label{eq.fdt1}
\end{equation}

Taking account of the multiple scattering between particle-hole pair within
the RPA, the Green's function is expressed as 
\begin{equation}
\hat{Y}^{{\rm T}}(q)= \hat{F}(q)[\hat{I}+\hat{\Gamma}\hat{F}(q)]^{-1}
 = \left[\hat{F}(q)^{-1}+\hat{\Gamma}\right]^{-1},
\label{eq.time_ladder}
\end{equation}
where
\begin{equation}
[\hat{\Gamma}]_{\xi_{2}\xi'_{2};\xi_{1}\xi'_{1}}=
\Gamma^{(0)}(\xi_{2}\xi'_{1};\xi_{1}\xi'_{2}), 
\end{equation}
and the particle-hole propagator $\hat{F}(q)$ is defined as
\begin{widetext}
\begin{equation}
[\hat{F}(q)]_{\xi_2\xi'_{2};\xi_1\xi'_{1}} \equiv
-i\frac{2}{N}\sum_{{\bf k}}\int\frac{{\rm d}k_{0}}{2\pi}
[\hat{G}({\bf k+q},k_{0}+\omega)]_{\xi_{2}\xi_{1}}
[\hat{G}({\bf k},k_{0})]_{\xi'_{1}\xi'_{2}}.
\end{equation}
By substituting Eq. (\ref{eq.Green}) into the single-particle 
Green's function, we get
\begin{eqnarray}
[\hat{F}(q)]_{\xi_2\xi'_{2};\xi_1\xi'_{1}} 
 & = & \frac{2}{N}\sum_{{\bf k}}\sum_{\ell,\ell '}
 U_{\xi_{2}\ell}({\bf k+q})U_{\xi_{1}\ell}^{*}({\bf k+q})
 U_{\xi'_{1}\ell '}({\bf k})U_{\xi'_{2}\ell '}^{*}({\bf k})
\nonumber \\
 & \times & \left[\frac{[1-n_{\ell}({\bf k+q})]n_{\ell '}({\bf k})}
 {\omega-E_{\ell}({\bf k+q})+E_{\ell '}({\bf k})+i\delta}
 -\frac{n_{\ell}({\bf k+q})[1-n_{\ell '}({\bf k})]}
 {\omega-E_{\ell}({\bf k+q})+E_{\ell '}({\bf k})-i\delta}\right].
 \label{eq.green_pair}
\end{eqnarray}

We need a special care for the bound states, which appear 
below the energy continuum as a pole in 
$\hat{Y}^{T}(q)$. For the bound state $\omega >0$, since
$\hat{F}(q)$ is a Hermite matrix, we could 
diagonalize $\hat{F}(q)^{-1}+\hat{\Gamma}$ by a unitary matrix.
Let an eigenvalue be zero at $\omega=\omega_{B}({\bf q})$ 
with the eigenvector $B_{\xi\xi'}({\bf q})$. We could expand 
$[\hat{Y}^{{\rm T}}(q)]_{\xi_1\xi'_{1};\xi\xi'}$ around 
$\omega=\omega_{B}({\bf q})$ as 
\begin{equation}
\left[\hat{Y}^{{\rm T}}(q)\right]_{\xi_1\xi'_{1};\xi\xi'}=
\frac{[\hat{C}({\bf q})]_{\xi_1\xi'_{1};\xi\xi'}}{\omega-\omega_{B}({\bf q})+i\delta},
\label{eq.bound2}
\end{equation}
where 
\begin{equation}
[\hat{C}({\bf q})]_{\xi_1\xi'_{1};\xi\xi'}=\frac{B_{\xi_1\xi'_{1}}({\bf q})
B_{\xi\xi'}^{*}({\bf q})}
{\sum_{\xi_2\xi'_{2}\xi_3\xi'_{3}}B_{\xi_3\xi'_{3}}^{*}({\bf q})
\frac{\partial [\hat{F}({\bf q},\omega_B({\bf q}))^{-1}]_{\xi_3\xi'_{3};\xi_2\xi'_{2}}}
     {\partial\omega}B_{\xi_2\xi'_{2}}({\bf q})}.
\label{eq.wt.c}
\end{equation}
\end{widetext}
The correlation function is evaluated by inserting (\ref{eq.bound2})
into the right hand side of Eq.~(\ref{eq.fdt1}), which results in 
\begin{equation}
 \hat{Y}^{+-}(q)=2\pi
\hat{C}({\bf q})\delta(\omega-\omega_{B}({\bf q})).
\label{eq.bound3}
\end{equation}
Finally, the contribution to the RIXS intensity from the bound state
is given by substituting Eq. (\ref{eq.bound3}) into 
Eq.~(\ref{eq.second2}).

\section{\label{sect.5}Numerical results for RIXS spectra}
\begin{figure}
\includegraphics[width=8.0cm]{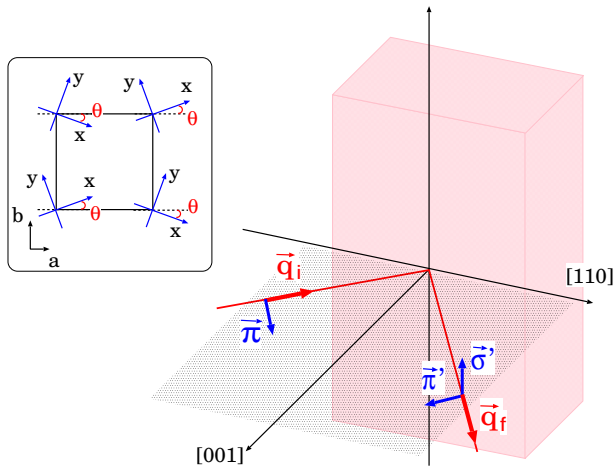}%
\caption{\label{fig.geometry} (Color online)
Geometry of 90$^{\circ}$ scattering. The scattering plane is perpendicular
to the $ab$ plane and intersects the plane along the $[110]$ direction.
The inset depicts the local coordinate frames, which are rotated by angle 
$\pm\theta$ around the $c$ axis.
}
\end{figure}

We consider the specific case of a 90$^{\circ}$ scattering angle
in accordance with the experiments.
The scattering plane is perpendicular to the IrO$_2$ plane, which intersects
the $ab$ plane with the $[110]$ direction, as illustrated
in Fig.~\ref{fig.geometry}. 
Since $\omega\sim 11.2$ keV and $|{\bf q}_i|\sim 5.7$ $\textrm{\AA}^{-1}$
at the Ir $L_3$ edge, only a few degrees of tilt of the scattering plane
could sweep the entire Brillouin zone. The local coordinate frame is defined by
rotating the $xy$ axes around the crystal $c$ axis with 
$\theta=11^{\circ}$ ($-11^{\circ}$) at A (B) sublattice, 
as shown in the inset of Fig. 2. Therefore the polarization vectors
are represented in the local coordinate frames as
\begin{eqnarray}
 \alpha_i &:& \pi = \left(\frac{\cos\theta\mp\sin\theta}{2},
                     \frac{\cos\theta\pm\sin\theta}{2},
                     \frac{1}{\sqrt{2}}\right),\\ 
 \alpha_f &:& \sigma' = \left(\frac{\cos\theta\pm\sin\theta}{\sqrt{2}},
                   \frac{-\cos\theta\pm\sin\theta}{\sqrt{2}},0\right),\\ 
          &:& \pi' = \left(-\frac{\cos\theta\mp\sin\theta}{2},
                      -\frac{\cos\theta\pm\sin\theta}{2},
                      \frac{1}{\sqrt{2}}\right),
\end{eqnarray}
where the upper and lower signs correspond to the A and B sublattices,
respectively. The incident x ray is assumed to have the 
$\pi$ polarization. 
Inserting these relations into Eq.~(\ref{eq.Mdef}), we obtain $\hat{M}$.

\subsection{Spectra for magnon}
We first study the magnetic excitations emerging as bound states in 
$Y^{T}(q)$, which may be called as magnons. 
We numerically evaluate $\hat{F}(q)$ by summing over ${\bf k}$ in 
Eq.~(\ref{eq.green_pair}) with dividing the first MBZ into 
$100\times 100$ meshes. The calculation is straightforward for $\omega$
below the energy continuum. The bound states are determined by adjusting 
$\omega$ to give zero eigenvalue in $\hat{F}(q)^{-1}+\hat{\Gamma}$.
In evaluating the corresponding intensity, we numerically carry out finite 
difference between $\omega=\omega_B({\bf q})$ and $\omega_B({\bf q})+0.01$eV
in Eq.~(\ref{eq.wt.c}) in place of the differentiation.

\begin{figure}
\includegraphics[width=8.0cm]{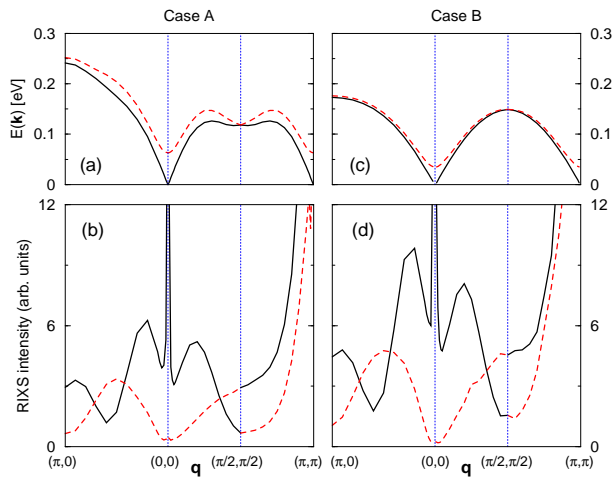}%
\caption{\label{fig.int}(Color Online)
Dispersion relation of magnon and corresponding RIXS 
intensity for ${\bf q}$ along symmetry directions. 
Panels (a) and (b) are for the parameters in Case A, while panels
(c) and (d) are in Case B. 
The (black) solid and (red) dotted lines 
are for the modes with lower and higher energies, respectively.
}
\end{figure}

Figure \ref{fig.int} shows the dispersion relation of magnon thus determined,
and the corresponding RIXS intensities at the $L_3$ edge. 
It is found that the magnon is split into two modes 
with slightly different energies, as already reported \cite{Igarashi2014-2}. 
Although such split modes are not confirmed, their dispersion relation is in
qualitative agreement with that derived from the RIXS experiment. In Case A,
the energies of magnon are given by $\omega_B(\pi,0)=0.241$ eV and $0.252$ eV 
at the $X$ point, while $\omega_B(\pi/2,\pi/2)=0.118$ eV at the $M$ point,
which slightly overestimate the magnon energies. On the other hand, 
in Case B,
we have $\omega_B(\pi,0)=0.173$ eV and $0.176$ eV at the $X$ point, and 
$\omega_B(\pi/2,\pi/2)=0.149$ eV at the $M$ point, which slightly underestimate
the magnon energies.   

The RIXS intensity at the $L_3$ edge is also shown in Fig. \ref{fig.int}.
Although the energy of magnon is periodic with the MBZ, the intensity is not 
periodic because of the presence of $v({\bf q})$ in Eq. (\ref{eq.Mdef}). 
In the narrow region around the $\Gamma$ point, the intensity of the 
mode with lower energy seems to diverge with $|{\bf q}|\to 0$. 
This arises from the staggered rotation of IrO$_6$ octahedra, and may 
be related to the presence of the weak ferromagnetism. The intensity of 
another mode ($\omega_B(0)=0.057$ eV in Case A and $0.035$ eV in Case B) 
is weak but finite. 
On the other hand, around the ${\bf q}=(\pi,\pi)$, 
the intensity of the mode with lower energy diverges with 
${\bf q}\to (\pi,\pi)$, due to a reflection of the antiferromagnetic order. 
The intensity of another mode with higher energy is finite but quite large. 
Although the magnon peak at the $X$ point has been interpreted as being 
separated into a one-magnon 
and a weak two-magnon peaks in the RIXS experiment (Fig. 4(c) in Ref.
\onlinecite{J.Kim2012}), it might 
be more appropriate to assign the two peaks as the split modes,
since the intensities of two-magnon excitations are expected to be
quite small. At the $M$ point, the separation of the intensity could not
be perceived, since the two modes are degenerate. 
It is found from the intensity curve that,
across the $M$ point, the wavefunction of the mode with low energy
is continuously connected to that with higher energy and vice versa. 
These characteristics mentioned above are consistent with 
the recent analysis on the basis of the 
localized spin model \cite{Igarashi2014-1}.
For the general values of ${\bf q}$, however, the intensity 
varies rather strongly with changing values of ${\bf q}$, 
in contrast with the monotonic change found in the localized spin model.

\subsection{Spectra for exciton} 
There emerge several bound states between the magnon modes and the continuous
states in the spectral function of the two-particle Green's function. 
The calculation of the bound states is the same as that of the magnon 
modes. For $\omega$ inside the energy continuum of electron-hole pair 
excitations,
we evaluate Eq.~(\ref{eq.green_pair}) by storing each 
$E_{\ell}({\bf k+q})-E_{\ell '}({\bf k})$ into segments with the width of 
$0.01$ eV for $100\times 100$ ${\bf k}$-points, resulting in the 
histogram representation of the imaginary part of $\hat{F}(q)$.
Setting $\omega$ at the center of each 
segment, we evaluate Eq.~(\ref{eq.green_pair}) and thereby 
Eq.~(\ref{eq.time_ladder}), and finally Eq.~(\ref{eq.second}). 

\begin{figure}
\includegraphics[width=8.0cm]{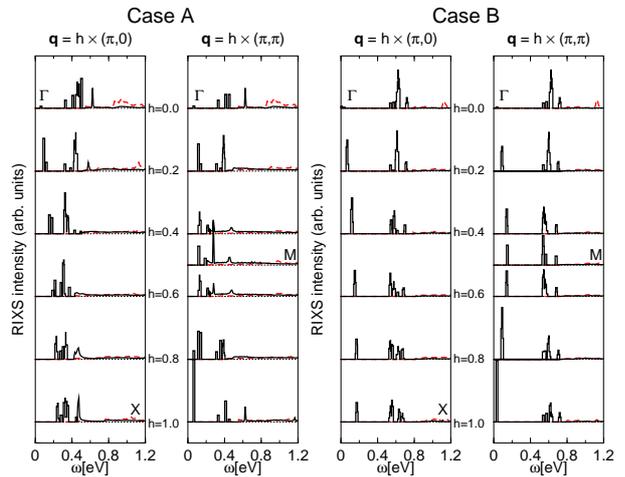}%
\caption{\label{fig.spectra}(Color Online)
The RIXS spectra as a function of energy 
loss $\omega$  for ${\bf q}$ along symmetry directions.
The (black) solid and (red) broken lines correspond to the 
intensities with and without 
taking account of the multiple scattering, respectively.
The $\delta$-function peaks are replaced by rectangles with their widths 
$0.02$ eV.
}
\end{figure}

Figure \ref{fig.spectra} shows the RIXS spectra thus evaluated as a function 
of $\omega$ for ${\bf q}$ along typical symmetry directions. 
The spectra are also shown without taking account of the multiple scattering 
($\hat{Y}^{T}(q)$ is replaced by $\hat{F}(q)$) for reference. 
The $\delta$-function peaks are replaced by rectangles with their widths 
$0.02$ eV. 
It is found that the bound states bear large part of exciton intensities. 
The exciton peaks are close to the magnon peaks around the $X$ and $M$ points 
in Case A, while they are well separated from the magnon peaks in Case B 
in agreement with the RIXS experiment. Since the bound states of 
excitons are composed mainly of the $j_{\rm eff}=1/2$ states of electron 
and $j_{\rm eff}=3/2$ 
states of hole, the larger value of $\zeta_{\rm SO}$ in Case B may lead to 
the larger separation between the exciton and magnon peaks. 
In the localized electron picture, the exciton 
peak is given by the excitation from the $j_{\rm eff}=3/2$ manifold 
to the $j_{\rm eff}=1/2$ manifold, where the dispersion is given by
the hopping in the antiferromagnetic isospin 
background \cite{Kim2012,Ament2011}.


\begin{figure}
\includegraphics[width=7.50cm]{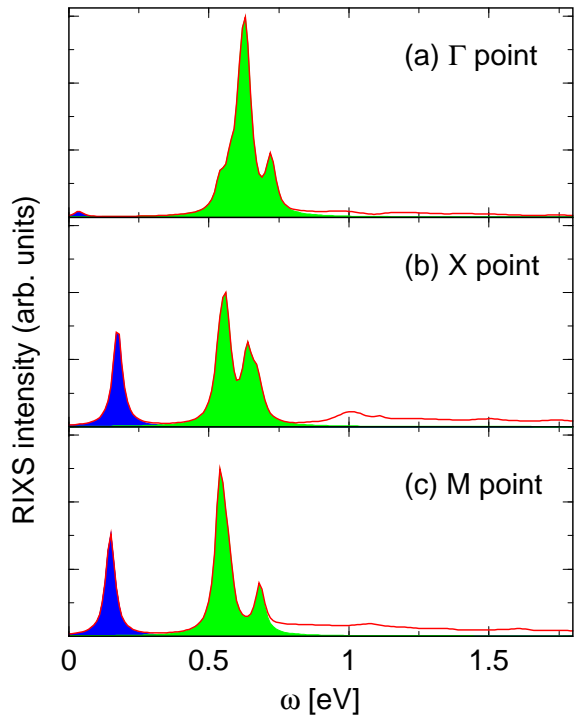}%
\caption{\label{fig.combined}(Color Online)
The RIXS spectra for both the magnon and exciton excitations 
as a function of energy loss $\omega$ at
(a) the $\Gamma$, (b) $X$, and (c) $M$ points, with $\zeta_{\rm SO}=0.45$ eV
and $t_2=0.25$ eV (Case B). 
The spectra are convoluted with Lorentzian function
with the full width half maximum 0.04 eV. 
The (blue) filled area around the peak with the lowest energy 
represents the magnon contribution.
The contribution from the bound state of exciton is
represented by the (green) filled area 
around $\omega \simeq 0.5-0.7$ eV.
The divergent intensity at $\omega=0$ are excluded at the $\Gamma$ point.
}
\end{figure}

For comparison with the experimental RIXS spectra, the calculated spectra
are convoluted with the Lorentzian function with the full width half maximum 
$0.04$ eV. Figure \ref{fig.combined} shows the result in Case B.
The peak with the lowest energy represents the magnon
contribution. The intensity diverging at $\omega=0$ is excluded at the 
$\Gamma$ point. The split of magnon modes could not be
distinguished at the $X$ point due to the convolution. 
The intensities of exciton peaks are two or three times larger than 
those of magnon modes, which ratio is comparable with the experiment
\cite{J.Kim2012}.

\section{\label{sect.6}Concluding remarks}

We have analyzed the $L$-edge RIXS spectra from Sr$_2$IrO$_4$ in an 
itinerant electron approach. Introducing a multi-orbital tight-binding model,
we have calculated the one-electron energy band within the HFA, 
and the Green's functions for particle-hole pair excitations 
within the RPA. 
The RIXS spectra have been evaluated from the Green's functions
within the FCA.
We have found two kinds of peaks in the RIXS spectra.

One is the peak of magnon, which arises from the bound state 
in $\hat{Y}^{T}(q)$. The dispersion of magnon is obtained in agreement 
with the experiment \cite{J.Kim2012}. 
We have predicted two-peak structures
with slightly different excitation energy $\sim 0.05$ eV due to the split
of magnon modes. Since the instrumental resolution is the same order of 
the split, it seems hard to detect the split in RIXS experiments. 
Some clue of the split, however, might be found with a careful examination 
of the spectral shape or by improving the experimental energy resolution.

Another is the peak of exciton, which also arises from the bound states in
$\hat{Y}^{T}(q)$. 
We have found large intensities concentrated on these peaks in comparison with
the intensities of continuous states.
The peak positions relative to the magnon peaks depend on parameters.
The larger value of $\zeta_{\rm SO}$ and the smaller value of $t_2$ (Case B)
seem to give the exciton peaks in better position in comparison with
the RIXS experiment. To make quantitative understanding of the spectra, 
however, it may be necessary to refine the present model by
including the hopping electron to further neighbors as well as the tetragonal 
crystal field or by including more correlation effects beyond the HFA and RPA. 
Studies along this direction are left in future.

\begin{acknowledgments}
This work was partially supported by a Grant-in-Aid for Scientific Research
from the Ministry of Education, Culture, Sports, Science, and Technology,
Japan. 
\end{acknowledgments}

\bibliographystyle{apsrev} 
\bibliography{paper3}

\begin{thebibliography}{40}
\bibitem[{Com()}]{Com3}
\bibinfo{note}{There exist several bound states between the magnon modes and
  the continuous states, which have been missed in Ref. [32].}
\end{thebibliography}

\end{document}